\documentclass[aps,twocolumn]{revtex4-1}
\pdfoutput=1 
\usepackage{graphicx}
\usepackage{bm}
\usepackage{braket}
\usepackage{pifont}
\usepackage{amsmath}
\usepackage{amsfonts}
\usepackage{amssymb}
\usepackage{graphicx}
\graphicspath{{./Figures/}}
\usepackage[FIGBOTCAP]{subfigure}
\usepackage{color,xcolor}
\usepackage{tensor}
\usepackage{natbib}

\begin{document}

\title{Quantum error-correction of continuous-variable states with realistic resources} 
\author{Josephine Dias}
\email{josephine.dias@uqconnect.edu.au}
\author{T.C. Ralph}

\affiliation{Centre for Quantum Computation and Communication Technology$,$ School of Mathematics and Physics$,$ University of Queensland$,$ Brisbane$,$ Queensland 4072$,$ Australia}

\date{\today}

\begin{abstract}
Gaussian noise induced by loss on Gaussian states may be corrected by distributing EPR entanglement through the loss channel, purifying the entanglement using a noiseless linear amplifier (NLA) and then using it for continuous-variable teleportation of the input state. Linear optical implementations of the NLA unavoidably introduce small amounts of excess noise and detection and source efficiency will be limited in current implementations. In this paper, we analyze the error-correction protocol with non-unit efficiency sources and detectors and show the excess noise may be partially compensated by adjusting the classical gain of the teleportation protocol.  We present a strong case for the potential of demonstrable error-correction with current technology. 
\end{abstract}

\maketitle

\section{Introduction}
Quantum communication enables promising new technologies like secure communication \cite{ gisin2002quantum,scarani2009security} and quantum teleportation \cite{pirandola2015advances}. While the advantages of these technologies are intriguing, quantum communication requires transmission of a quantum state over long distance. Losses in fiber have proven to severely limit the distance achieved with quantum communication protocols \cite{sangouard2011quantum}. It is therefore, of paramount importance to be able to correct against the effects of loss and decoherence on the channel along which the delicate quantum states are sent. 

Quantum information protocols may be grouped into two distinct regimes, discrete variable (DV) where information is encoded in a finite dimensional basis, for example the polarization of single photons \cite{nielsen2002quantum}, and the continuous-variable (CV) regime where information is encoded in an infinite dimensional basis and measurements are made using homodyne and heterodyne detection \cite{weedbrook2012gaussian}. Due to the ease of generation, manipulation and detection of the Gaussian states required for CV protocols, it promises simple and efficient implementations of quantum information protocols \cite{munro2015inside}. 

It is known that error-correction of Gaussian noise on Gaussian states using only Gaussian resources is impossible \cite{niset2009no}. Accordingly, to correct against Gaussian noise on Gaussian states, a non-Gaussian resource is required. This was exemplified in \cite{ralph2011quantum} where a protocol to correct against Gaussian noise induced by loss on Gaussian states was presented. This protocol works by using the noiseless linear amplifier (NLA) \cite{ralph2009nondeterministic} to purify entanglement distributed through the loss channel. The purified entanglement is then used for CV teleportation of the input state. 

In \cite{dias2017quantum}, it was shown that linear optical implementations of the NLA result in added noise to the error-corrected output state of the protocol. In the following, we present a way of partially compensating for the added noise by optimizing the classical gain of the CV teleportation. We particularly wish to determine if current source and detector efficiencies are sufficient to demonstrate quantum error-correction via this protocol. We answer this question in the affirmative.

The paper is arranged in the following way, in Sec.~\ref{sec:ec} we review the error-correction protocol from Ref.~\cite{ralph2011quantum}, in Sec.~\ref{sec:g} we introduce a method to improve the protocol by optimizing the classical teleportation gain and finally in Sec.~\ref{sec:exp} we model the protocol under realistic conditions of non-unit efficiency sources and detectors to determine the level of error-correction that may be demonstrated with current technology. 
%

\section{The Error-Correction Protocol\label{sec:ec}}
\begin{figure}[h!]
\centering
\subfigure[]{\label{fig:loss-channel}\includegraphics[width=\linewidth]{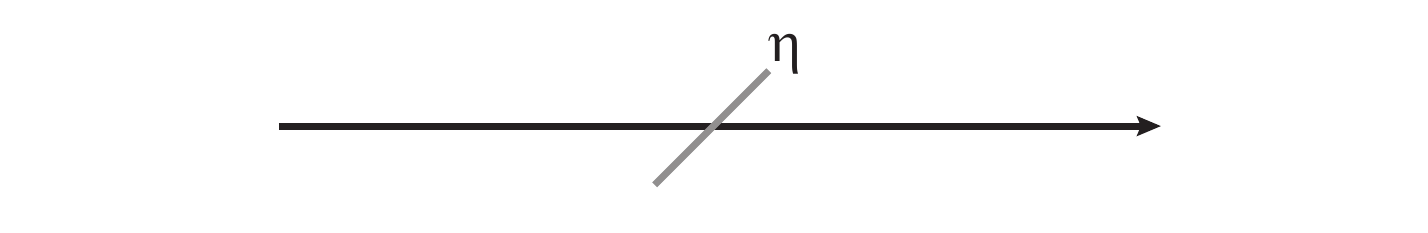}}
\subfigure[]{\label{fig:error-correction1}\includegraphics[width=\linewidth]{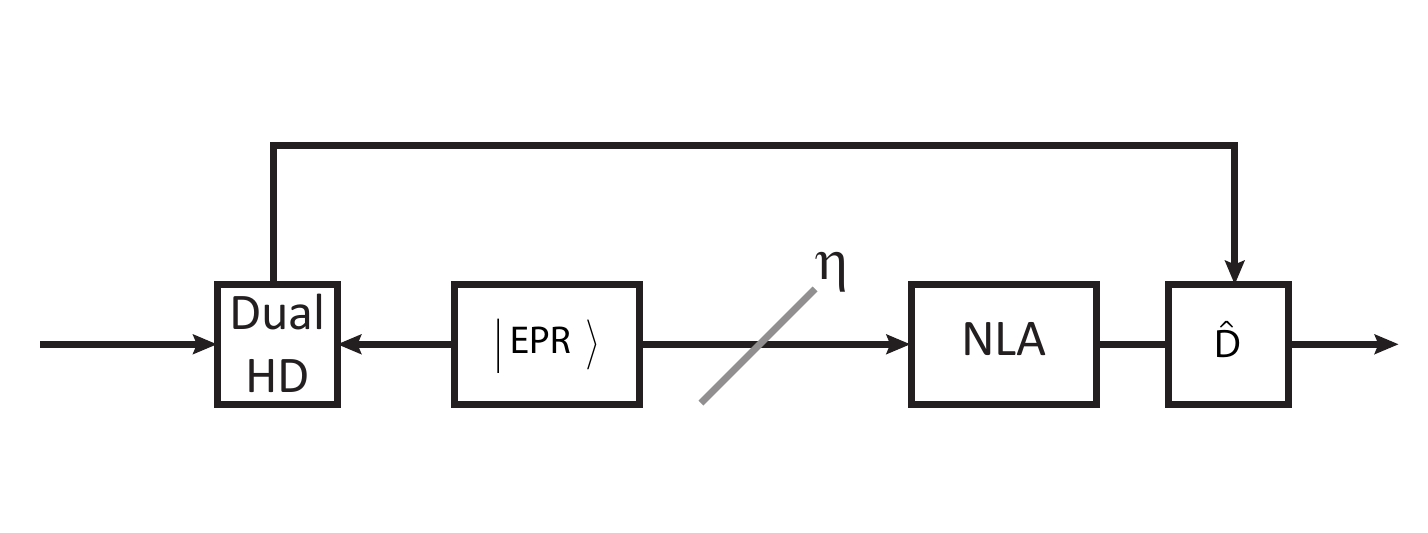}}
\caption{\subref{fig:loss-channel} A lossy channel of transmission \(\eta\) takes an input coherent state \(\ket{\alpha}\) to \(\ket{\sqrt{\eta} \alpha}\).  \subref{fig:error-correction1} Protocol for quantum error-correction of CV states. EPR entanglement is distributed through the lossy channel. The NLA distills the entanglement which is then used for CV teleportation of the input state. Ideally, this protocol takes an input coherent state \(\ket{\alpha}\) to output state \(\ket{g\sqrt{\eta}\chi\alpha}\) \cite{ralph2011quantum}.}
\label{fig:error-correction}
\end{figure}

The error-correction protocol is pictured in Fig.\ref{fig:error-correction1}. To correct against loss on Gaussian states, a two-mode Gaussian squeezed state, otherwise known as Einstein-Podolsky-Rosen (EPR) state, is generated and one arm is sent through the loss channel.  The EPR state has been degraded by the loss and is then distilled to increase the strength of the entanglement. This is achieved with the noiseless linear amplifier \cite{ralph2009nondeterministic} which forms the non-Gaussian resource required in this error-correction protocol. The NLA is a non-deterministic operation, with success probability decreasing as gain of the NLA increases. When the NLA has heralded successful operation, the distilled EPR state is then used for CV teleportation. The input state to be error-corrected is mixed with the other arm of the EPR state on a 50:50 beam splitter. Dual homodyne detection is then performed and the results are sent using a classical signal to the end of the channel. A displacement is then conducted on the output state depending on the outcome of the measurement and a classical gain \cite{ralph1999characterizing}. 

A coherent state \(\ket{\alpha}\) passing through a loss channel of transmission \(\eta\) becomes transformed as
\begin{equation}
\ket{\alpha}\to\ket{\sqrt{\eta}\alpha} .
\label{eq:transform loss}
\end{equation}
We compare this to the same state passing through the error-correction protocol which performs the following transformation 
\begin{equation}
\ket{\alpha}\to\ket{g\sqrt{\eta}\chi\alpha}.
\label{eq:transform}
\end{equation}
The effective transmission \(\eta_{\mathrm{eff}}\) of the error-corrected channel is 
\begin{equation}
\eta_{\mathrm{eff}}=g^2 \eta \chi^2 ,
\end{equation}
where \(g\) is the amplitude gain of the NLA, and \(\chi\) is the entanglement strength of the EPR (or two-mode squeezed) state. Therefore, we may introduce a condition for error-correction, that the effective transmission \(\eta_{\mathrm{eff}}\) is greater than the initial transmission of the loss channel. This condition simplifies to
\begin{equation}
g_{\mathrm{min}}>\frac{1}{\chi} . \label{eq:gmin}
\end{equation}

\begin{figure}[h]
\centering
\includegraphics[width=0.99\linewidth]{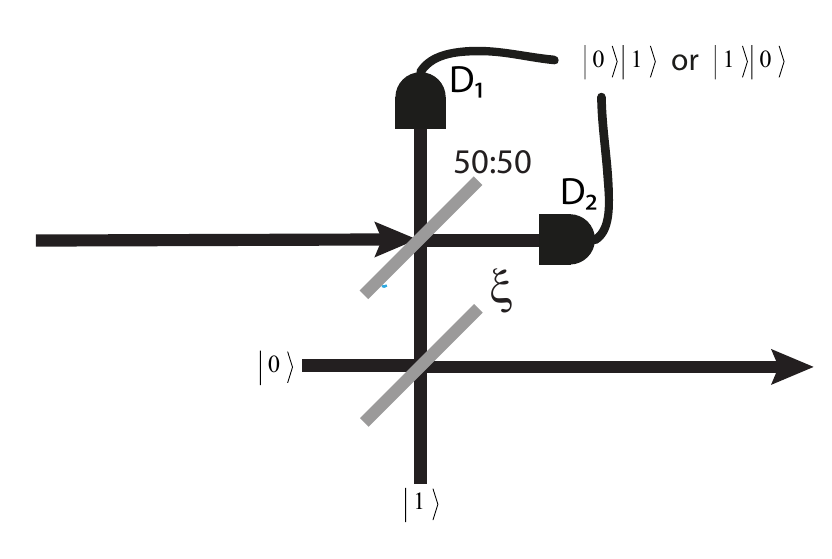}
\caption{NLA with a single quantum scissor.
Successful operation is heralded when a single photon is detected at \(D1\) and none at \(D2\) or vice versa. The gain  \(g\) is controlled by the tunable beam splitter ratio \(\xi\), related to the gain by  \(g = \sqrt{\left(1-\xi\right) /\xi}\). \label{fig:QS1_NLA}}
\end{figure}

The required transformation \eqref{eq:transform} is successfully achieved for an ideal, unphysical NLA. To illustrate what this means consider the physical linear optical implementation of the NLA. An input state to be amplified is split evenly into \(N\) modes and each mode is then passed through a single modified quantum scissor (QS) device \cite{pegg1998optical} (shown in Fig.~\ref{fig:QS1_NLA}). When each quantum scissor is successful, the outputs of the quantum scissors are coherently recombined to form the amplified output state. The correct transformation \(\ket{\alpha}\to\ket{g \alpha}\) is only achieved as the number of quantum scissors approaches infinity (\(N\to \infty\)). While the correct transformation is approximately achieved for large \(N\), this comes at the expense of a complex experimental set-up and a reduced success probability. This is because success probability of the NLA suffers an exponential decrease with \(N\). 

For the rest of this paper, we present results for the simplest case of the NLA, a single quantum scissor (\(N=1\)). The reason for this is two-fold, it represents the easiest experimental implementation and achieves maximal probability of successful operation. 
 
 The single quantum scissor NLA performs the  transformation
\begin{equation}
\hat{T}_1(\alpha\ket{0} +\beta\ket{1}+ \gamma\ket{2}+...) = \sqrt{\frac{1}{g^2+1}}(\alpha\ket{0} + g\beta\ket{1}) .
\label{eq:single_qs}
\end{equation}
Any higher order terms in the input state will be truncated with this operation. As noted in \cite{dias2017quantum}, this truncation operation results in a small amount of excess noise being introduced to the output state of the protocol. 

As the effect of this truncation on large amplitude input states is severe, we can expect a larger amount of excess truncation noise to be introduced for a larger input state to the NLA. For this reason, this error-correction protocol works best in the high-loss regime, i.e. where the state has been heavily attenuated from loss before passing through the NLA \cite{ralph2011quantum,dias2017quantum}.

Additionally, due to the presence of the truncation noise we require a new measure to quantify whether the channel has been error-corrected. The minimum gain condition \eqref{eq:gmin} only considers the effective transmission of the channel, however with excess noise this is not sufficient to quantify channel improvement. To evaluate the performance of the error-correction protocol we compute the Gaussian entanglement of formation (GEOF). We use the Gaussian entanglement of formation as it satisfies Gaussian extremality \cite{wolf2006extremality}, i.e. the GEOF of a non-Gaussian state with the same covariance matrix is a lower bound for the entanglement of formation. This is important as the output state from the error-correction protocol is non-Gaussian, therefore results presented in this paper may underestimate entanglement, but will not overestimate the amount of entanglement.

To evaluate whether our protocol has been effective at improving the channel, we compute the GEOF of a two-mode squeezed state with one arm of the entangled state distributed through the error-correction protocol.  We compare that to the entanglement of formation of the same state distributed through the same initial loss without error-correction. When the GEOF of the error-corrected channel surpasses that of the loss channel, we know the channel has been improved and this is the minimum requirement for error-correction. The GEOF is given by
\begin{equation}
\mathcal{E} = \cosh^2 r_0 \log_2 \left(\cosh^2 r_0\right)-\sinh^2 r_0 \log_2 \left(\sinh^2 r_0\right) .
\end{equation}
For the case of the uncorrected channel, the parameter \(r_0\) is
\begin{equation}
r_0 = \frac{1}{2} \ln \left[ \frac{1+\zeta\sqrt{\eta}}{1-\zeta\sqrt{\eta}} \right] 
\label{eq:r0}
\end{equation}
for finite squeezing \(\zeta\) and loss on one arm of the two-mode squeezed state \(\eta\) \cite{tserkis2017quantifying}. For the case of the error-corrected channel, the covariance matrix of the Gaussian approximated output state was calculated from the first and second moments of the output state (calculations shown in the Appendix) and the GEOF was calculated following Ref.~\cite{tserkis2017quantifying}.
 \section{Optimizing the teleportation gain \label{sec:g}}
In \cite{dias2017quantum}, this error-correction protocol was used to construct a continuous-variable quantum repeater. However, building truncation noise represented one of the most prominent limitations with this protocol.  In this section, we present a method for improving the outcome of the error correction protocol in the presence of this truncation noise. 

As part of the CV teleportation scheme, a dual homodyne measurement is performed on the mix of the input state and the EPR state. The results of this measurement are then sent via a classical signal to the end of the channel where a displacement is made according to the results of the measurement. The displacement depends not only on the results of the measurement, but also on a classical gain denoted here as \(\lambda\). In the presence of loss only, it is known that the optimal gain \(\lambda\) depends on the strength of the squeezing of the EPR source and  the effective transmission of the channel. In the case for our error-correction protocol with an ideal, unphysical NLA, this quantity would be \(\lambda= g\sqrt{\eta}\chi \) for optimal performance of the CV teleporter. 
\begin{figure}[h!]
\centering
\subfigure[]{\label{fig:GainTuned}
\includegraphics[width=0.85\linewidth]{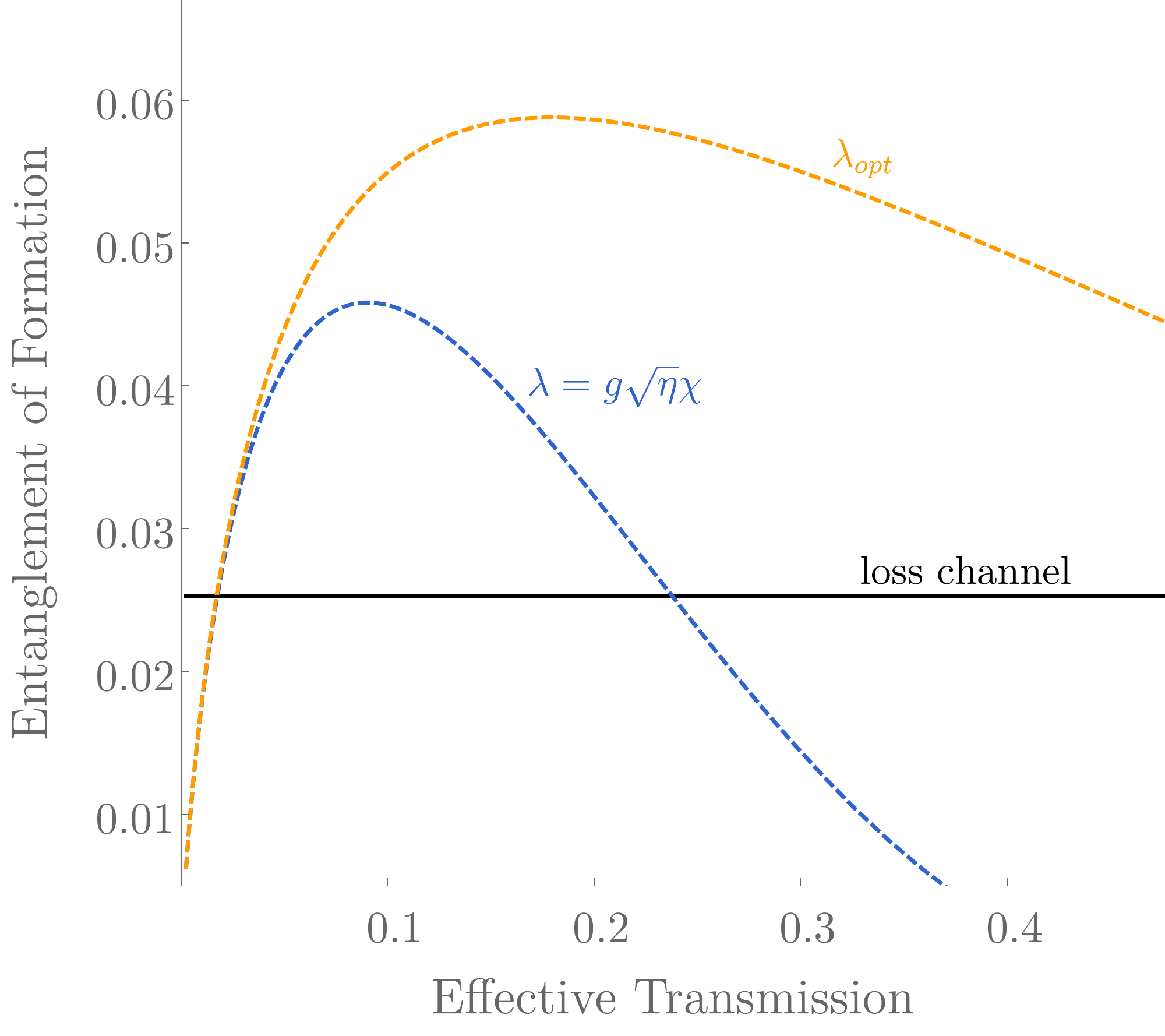}}
\subfigure[]{\label{fig:GainTunedPSuccess}\includegraphics[width=0.85\linewidth]{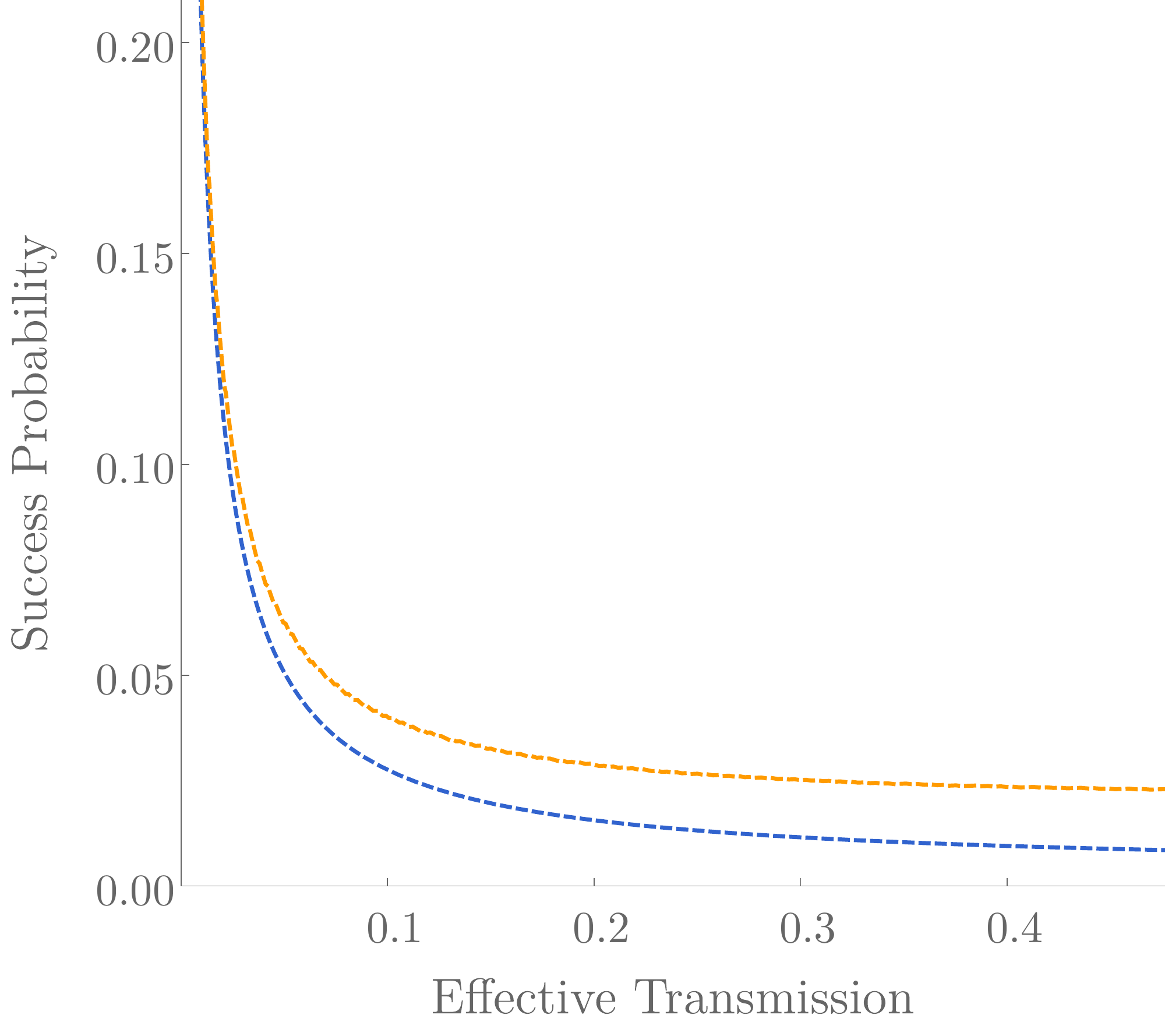}}
\caption{Improving the protocol by adjusting the classical teleportation gain. \subref{fig:GainTuned} Entanglement of formation of a Gaussian two-mode squeezed state (EPR strength \(\zeta=0.5\)) where one arm is distributed through a loss channel of transmission \(\eta=0.01\) (black, solid line).  EPR strength in the CV teleporter has been set to \(\chi=0.5\) (corresponding to \(r\approx0.55 \) where \(\chi=\tanh r\)). The blue, dashed line represents the error-correction protocol operating with the classical gain scaling factor \(\lambda=g\sqrt{\eta}\chi\). The yellow line uses a numerically optimised teleportation gain \(\lambda_{opt}\) to produce higher entanglement of formation and error correction over a larger range of effective transmission. \subref{fig:GainTunedPSuccess} The success probability for the same case as in \subref{fig:GainTuned}. By optimising \(\lambda\), the NLA gain may be reduced resulting in an improvement in the success probability.}\label{fig:GainTunedResults} 
\end{figure}
However, this known classical gain for the loss only case is not optimal in the presence of excess truncation noise. By slightly increasing \(\lambda\), we may observe a better outcome. 
 In Fig.~\ref{fig:GainTuned}, we see that the range of effective transmission achieving demonstrable error-correction has increased. We refer here to the region where the entanglement of formation of the error-correction protocol (dashed lines) is above that of the initial loss channel (black, solid line).  For the same parameters, the success probability of the NLA is shown in Fig.~\ref{fig:GainTunedPSuccess}. An increase in success probability and entanglement of formation for the optimised \(\lambda_{opt}\) means the error-correction protocol has been unambiguously improved, both in entanglement capacity through the channel and efficiency. 

Qualitatively, an increase in \(\lambda\) results in an increase in effective transmission and an increase in added noise from the CV teleportation protocol. As a result, there is an optimal value for the classical gain \(\lambda\) depending on the squeezing in the CV teleporter \(\chi\), the loss on the channel \(\eta\) and the gain of the NLA \(g\). For the results in this paper, \(\lambda\) was optimised numerically to produce the highest entanglement of formation of the protocol.

\section{Modeling experimental inefficiencies
 \label{sec:exp}}
 With our now improved outlook on the protocol due to the tuning of the CV teleportation gain, we now ask whether current technology permits demonstrable channel improvement using this protocol. Thus far, specific elements of the error-correction protocol have been experimentally implemented with promising results. In \cite{furusawa1998unconditional}, CV teleportation was experimentally implemented for the first time following the proposal in \cite{braunstein1998teleportation}. Since then, teleportation fidelities up to 83\% have been reported \cite{yukawa2008high}. 
 
 In  \cite{xiang2010heralded,ferreyrol2010implementation}, physical implementation of the NLA was realized with a single quantum scissor. As noted in \cite{kocsis2013heralded}, experimental imperfections in the NLA impact as follows: detection inefficiency reduces the probability of successful operation, inefficiency in the single photon source and lack of photon number resolution causes a gain saturation effect. 

\begin{figure}[h!]
\centering
\subfigure[]{\label{fig:EOF1}
\includegraphics[width=0.85\linewidth]{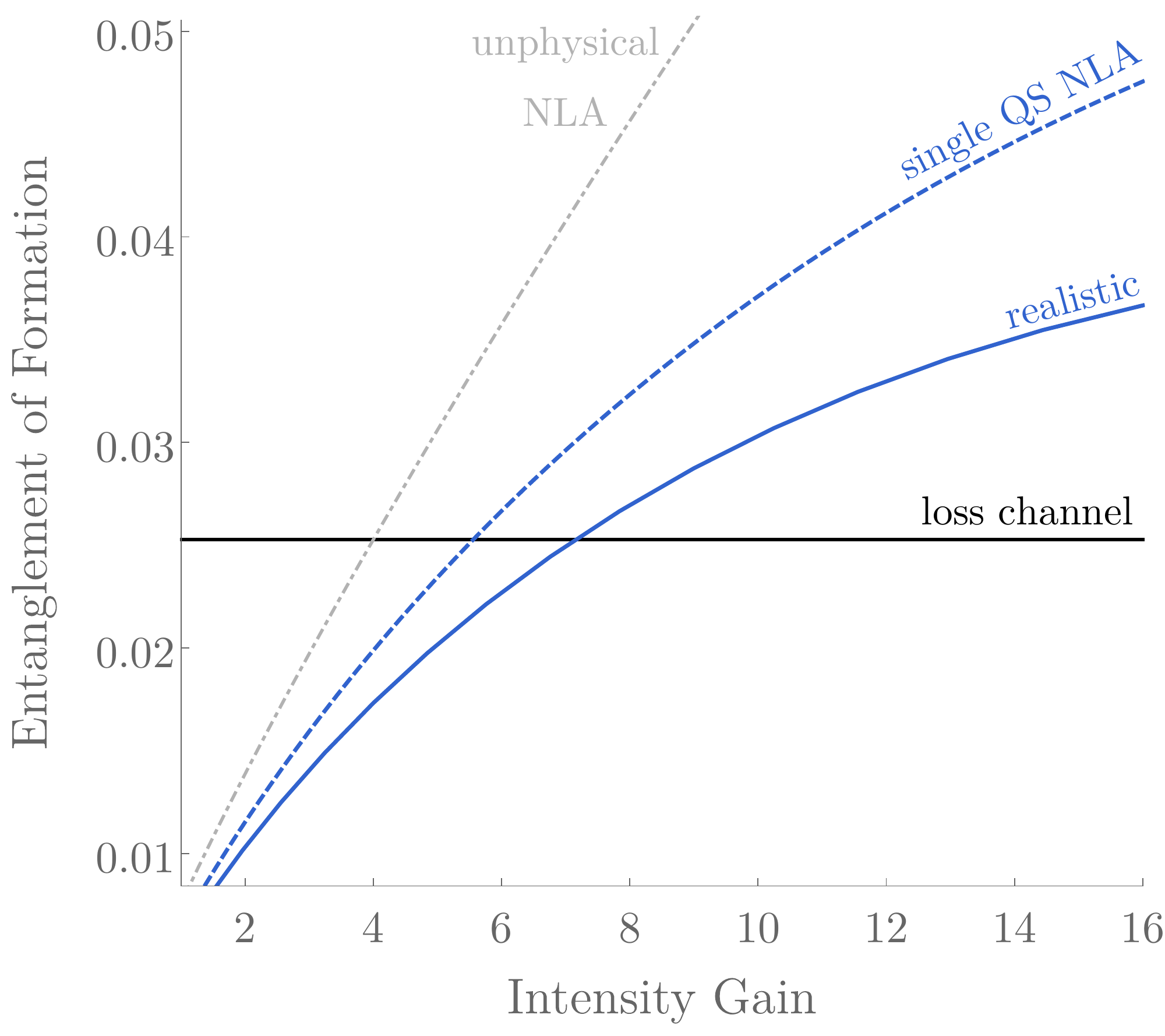}}
\subfigure[]{\label{fig:PSuc1}\includegraphics[width=0.85\linewidth]{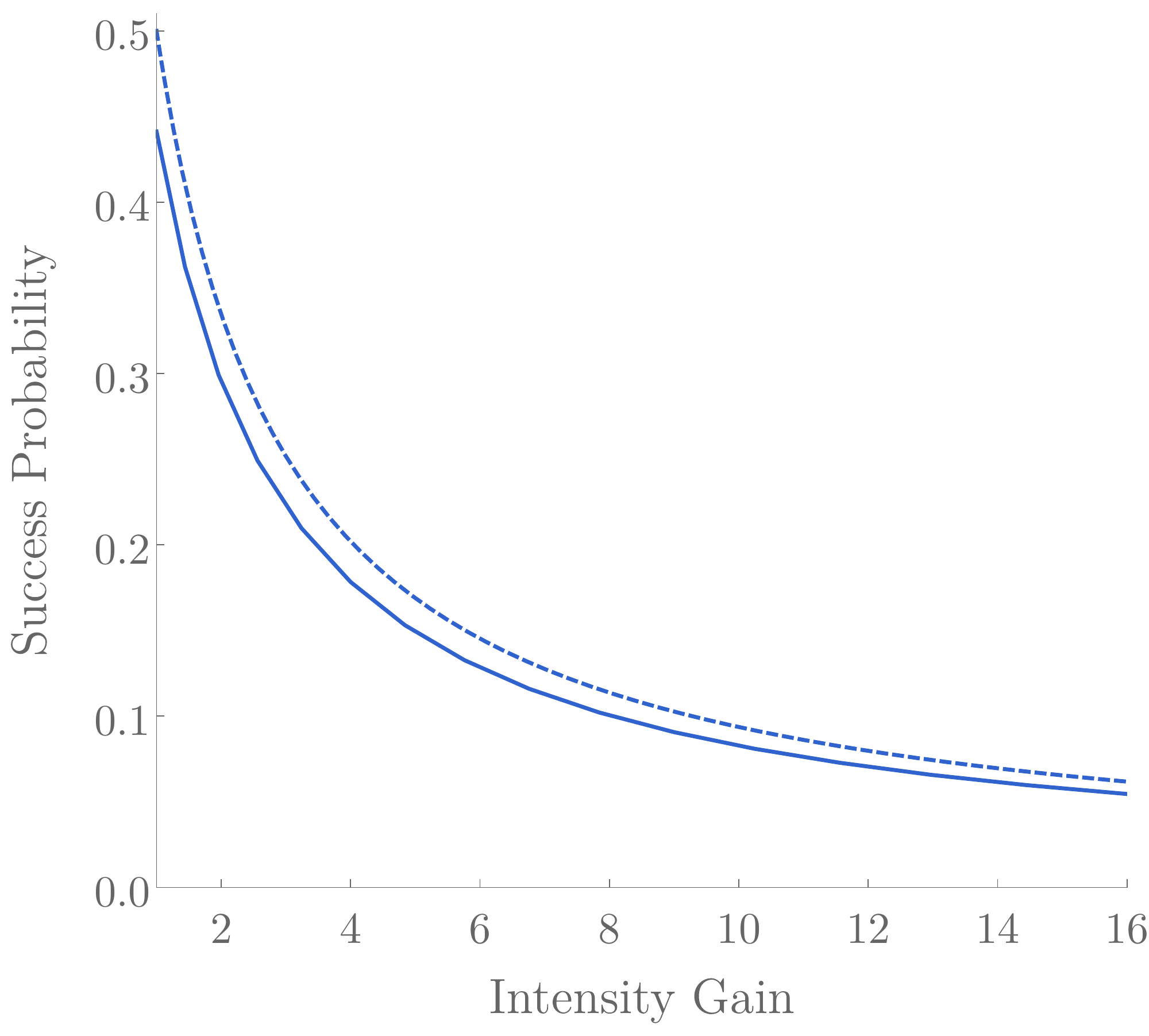}}
\caption{The error-correction protocol for an initial loss channel of transmission \(\eta=0.01\). \subref{fig:EOF1} The entanglement of formation of a Gaussian two-mode squeezed state (EPR strength \(\zeta=0.5\)) where one arm is distributed through loss \(\eta\) (black, solid line). The grey, dot-dashed line represents error-correction with the NLA operating in the unphysical limit of \(N\to\infty\), and passes the black line at \(g_{\mathrm{min}}>1/\chi\) \eqref{eq:gmin}. The dashed dark blue line is the EOF of the same state distributed through the error-correction protocol with a single quantum scissor NLA (with perfect sources and detectors).  EPR strength in the CV teleporter has been set to \(\chi=0.5\). The solid dark blue line includes realistic non-unit efficiency sources and detectors, using homodyne detection efficiencies of \(\tau=0.98\), single photon source efficiency of \(\epsilon=0.7\) and single photon detection efficiency \(\delta=0.9\). \subref{fig:PSuc1} The probability of successful operation of the NLA with the same parameters as Fig.~\ref{fig:EOF1}.}
\label{fig:results1}
\end{figure}
\begin{figure}[h!]
\centering
\subfigure[]{\label{fig:EOF2}
\includegraphics[width=0.85\linewidth]{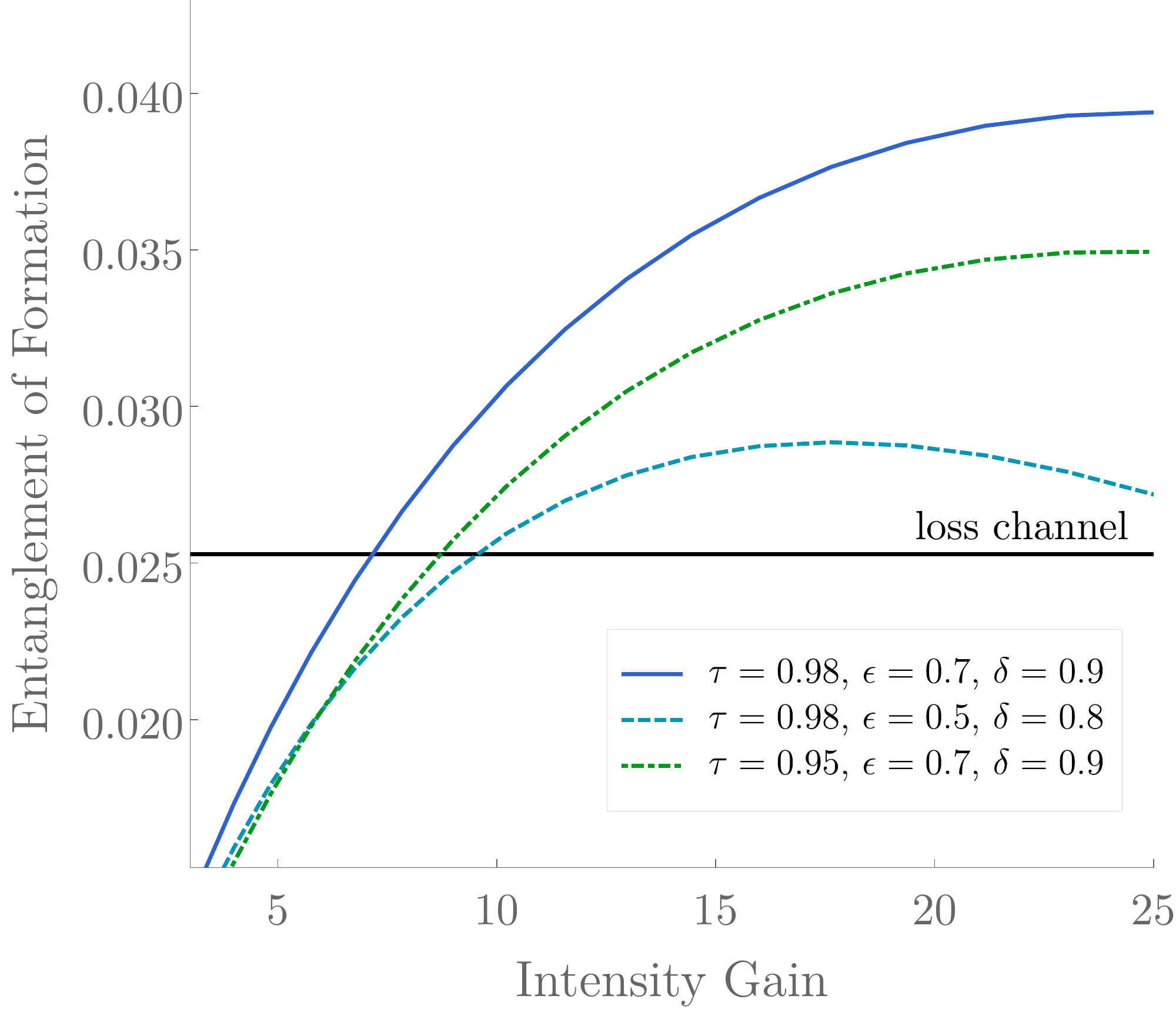}}
\subfigure[]{\label{fig:PSuc2}\includegraphics[width=0.85\linewidth]{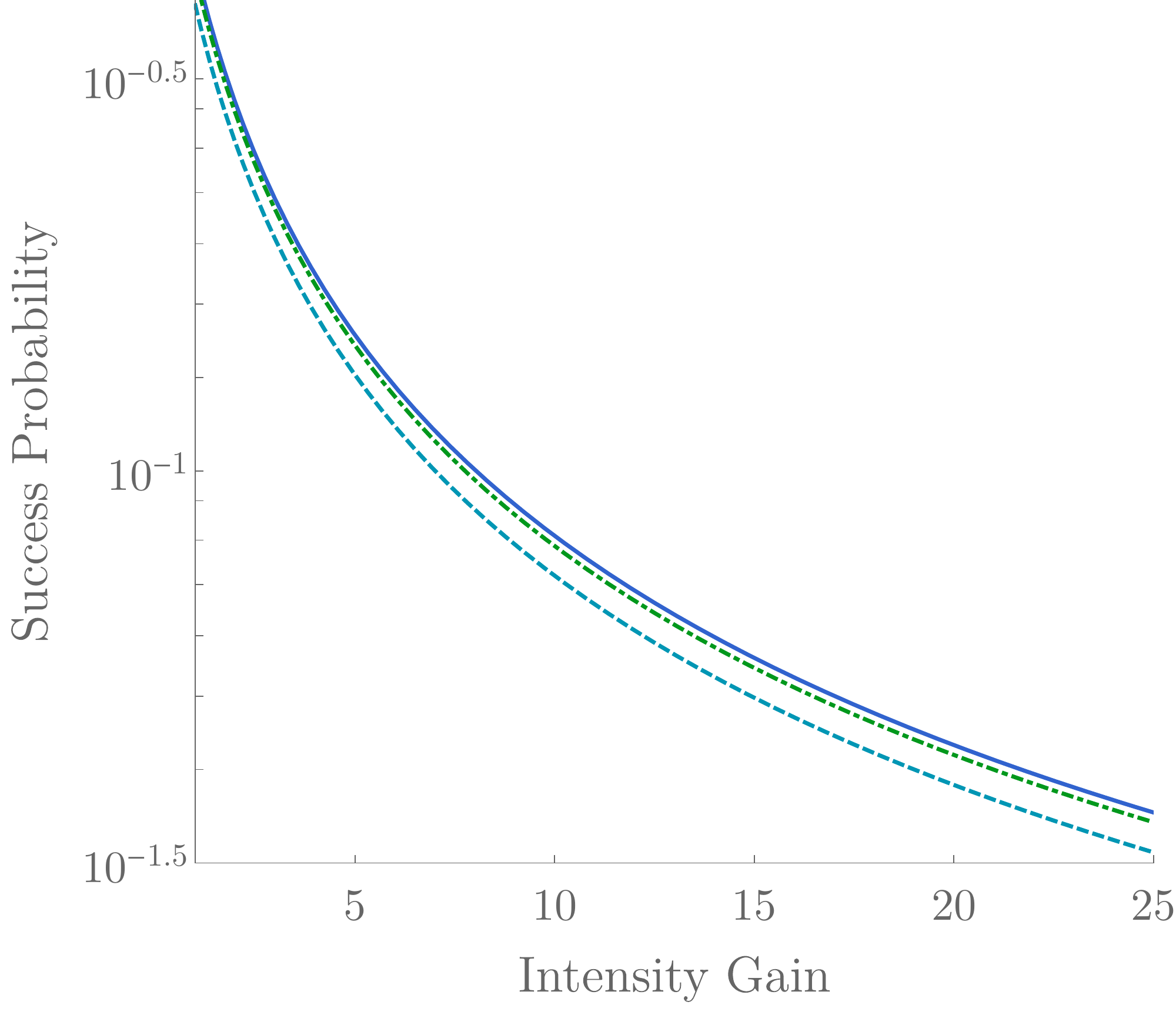}}
\caption{The effect of reduced source and detection efficiencies. \subref{fig:EOF2} The entanglement of formation of a Gaussian two-mode squeezed state (EPR strength \(\zeta=0.5\)) where one arm is distributed through a loss channel of transmission \(\eta=0.01\) (black, solid line). The solid, dark blue line corresponds to  that of the same state distributed through the error-correction protocol with the same parameters as in Fig.~\ref{fig:results1}. The light blue, dashed line has the same homodyne efficiency, but single photon elements in the NLA have reduced efficiency. The green, dot-dashed line has reduced homodyne efficiency but maintains the same NLA efficiency of the dark blue line. \subref{fig:PSuc2} Log plot of the probability of successful operation of the NLA with the same parameters as Fig.~\ref{fig:EOF2}. }
\label{fig:results2}
\end{figure}

\begin{figure}[h!]
\centering
\subfigure[]{\label{fig:EOF3}
\includegraphics[width=0.85\linewidth]{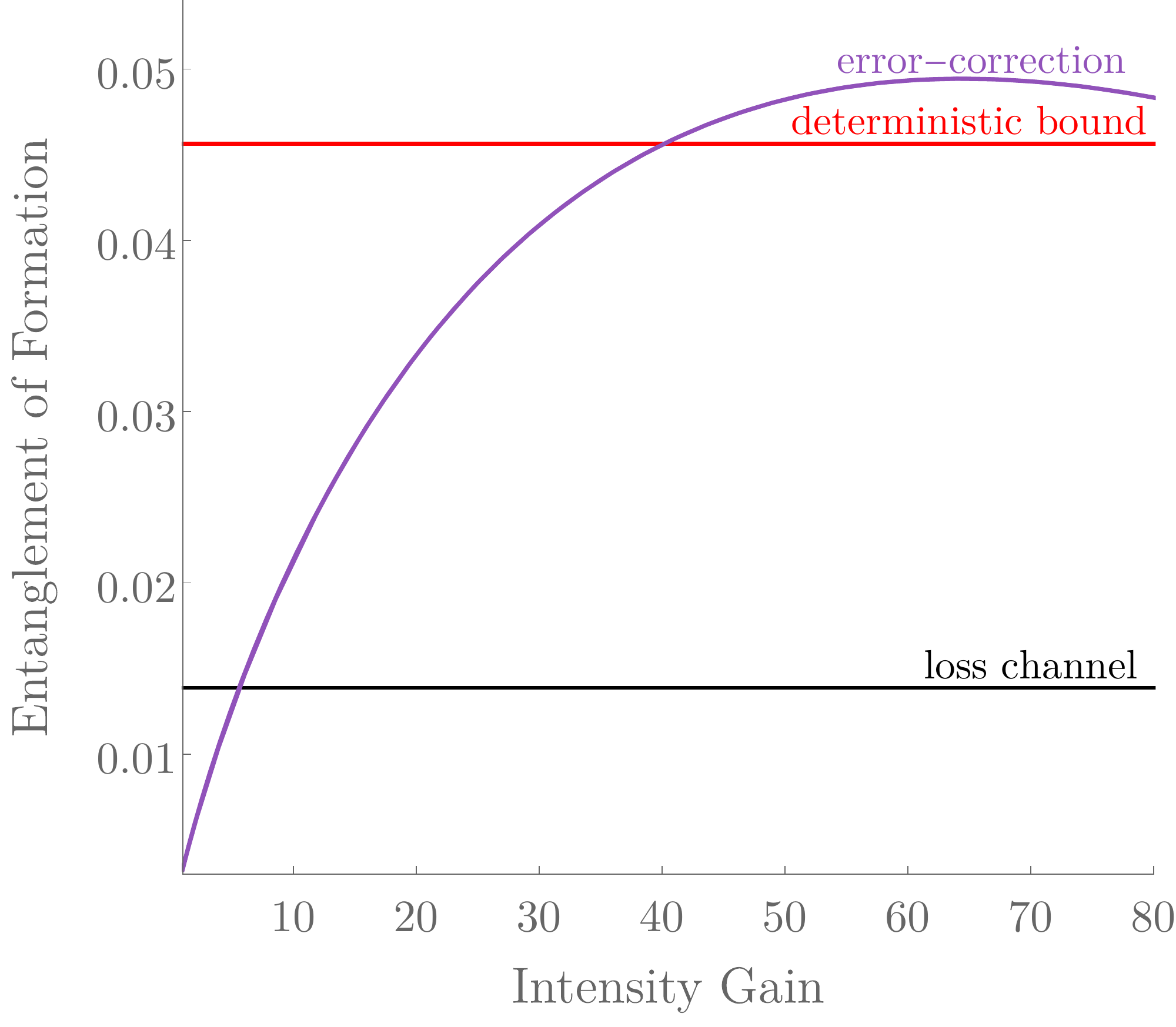}}
\subfigure[]{\label{fig:PSuc3}\includegraphics[width=0.85\linewidth]{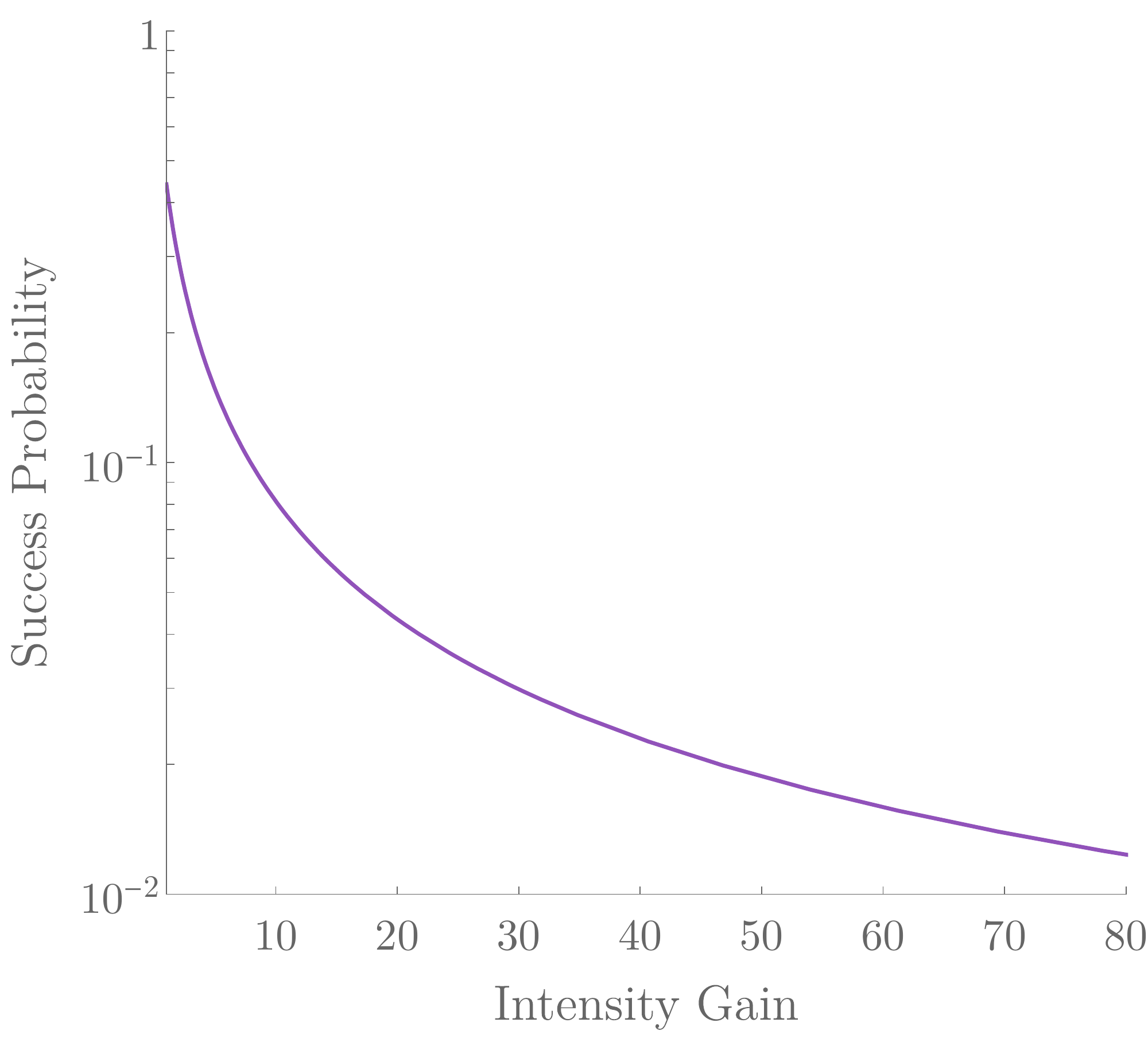}}
\caption{An example of the parameters required to beat the deterministic bound of entanglement. These results use an initial loss channel of transmission \(\eta=0.005\). EPR strength in the CV teleporter has been set to \(\chi=0.5\). \subref{fig:EOF3}   The entanglement of formation of a Gaussian two-mode squeezed state (EPR strength \(\zeta=0.5\)) where one arm is distributed through loss \(\eta\) (black, solid line). The deterministic bound representing the maximum entanglement achieved by an infinitely squeezed state passing through the same loss channel \(\eta\) (red, solid line).  The solid purple line corresponds to the error-correction protocol with homodyne detection efficiencies of \(\tau=0.98\), single photon source and detection efficiency has been set to \(\epsilon=\delta=0.9\).  \subref{fig:PSuc3} Log plot of the probability of successful operation of the NLA with the same parameters as Fig.~\ref{fig:EOF3}.}
\label{fig:results3}
\end{figure}

Notably, in \cite{ulanov2015undoing} the NLA was used to distil EPR entanglement that had been degraded by loss. In this experiment, by using the NLA, EPR entanglement was recovered to the original strength after it had passed through a loss channel of transmission \(\eta=0.05\).  

In the aforementioned implementations, Kocsis et al. report maximum achieved intensity gains of \(g^2= 5.7\pm0.5\) \cite{kocsis2013heralded} and Ulanov et al. report gains of \(g^2=10-12\) \cite{ulanov2015undoing}. Naturally, we ask whether the error-correction protocol can demonstrate channel improvement in the regime of these physically realizable gains. To address this problem, we have modeled the error-correction protocol following the approach of \cite{dias2017quantum} but now including the noise due to losses in homodyne detection and losses in the single photon source and detector within the NLA (details of the calculations have been included in the appendix). 

Fig.~\ref{fig:EOF1} shows that the unphysical NLA using \(N\to \infty\) quantum scissors  achieves an error-corrected channel for the gain condition \eqref{eq:gmin} when the entanglement of formation surpasses that of the original loss channel. When the protocol is modeled using the single quantum scissor NLA, the gain required for channel improvement is slightly higher due to truncation noise. Furthermore, under realistic conditions of imperfect sources and detectors in the homodyne measurement and NLA we find that the gain required for channel improvement is again increased. The solid blue line in Fig.~\ref{fig:results1} models the case for homodyne efficiencies of \(\tau=0.98\), single photon source  efficiency of \(\epsilon=0.7\) and single photon detection efficiency \(\delta=0.9\). For these specific parameters, NLA intensity gains of \(g^2>7.1\) may achieve a demonstrable channel improvement.  

We then ask how detrimental less efficient sources and detectors are to the outcome of the error-correction protocol. In Fig.~\ref{fig:results2}, we compare the result from Fig.~\ref{fig:results1} to the case where single photon source efficiency has been reduced to \(\epsilon=0.5\) and single photon detection efficiency reduced to \(\delta=0.8\). We also compare this to the case where the single photon element efficiency has remained the same, but homodyne efficiency has been reduced to \(\tau=0.95\). As expected, the gain required for error-correction has been slightly increased with the decrease in source or detection efficiency. However, the gains required for error-correction remains in the regime of previously physically realized NLA gains. 

As was also noted in \cite{dias2017quantum}, due to the truncation noise incurred by the single quantum scissor implementation of the NLA, smaller amplitude input states to the NLA are preferred. As such the protocol performs best in the high loss regime where the EPR entanglement is attenuated significantly, before being purified with the NLA. This is the reason both Figs.~\ref{fig:results1} and \ref{fig:results2} use the very high loss regime to demonstrate channel improvement (\(\eta=0.01\)). 

Finally, we ask if this protocol, despite the noise due to truncation and experimental inefficiencies can surpass the deterministic bound for entanglement. The deterministic bound represents the maximum entanglement of formation achieved by passing an unphysical infinitely squeezed state through the same loss channel. This bound is given by Equation~\ref{eq:r0} with \(\zeta\to 1\). To address this question, we present the results in Fig.~\ref{fig:EOF3} where the protocol has been shown to produce high enough entanglement of formation to surpass the deterministic bound. This was achieved by reducing the transmission of the loss channel to \(\eta=0.005\) as well as increasing the single photon source efficiency to \(\epsilon=0.9\). Given these parameters, gains of \(g^2 >41 \) are required for the entanglement of formation to surpass the deterministic entanglement bound. 

For each of these cases we present the probability of successful operation of the NLA in Figs.~\ref{fig:PSuc1}, \ref{fig:PSuc2} and \ref{fig:PSuc3}. Note that this is the probability of successfully creating the distilled entangled resource state with the NLA. Once created, the input state may be teleported deterministically. For each of these cases, a minor decrease can be seen in the probability of success caused by imperfect single photon detectors (\(\delta<1\)) and homodyne detectors (\(\tau<1\)). 
\section{Conclusion}

We have shown here that by adjusting the classical gain of the CV teleporter, the outcome of this error-correction protocol may be significantly improved in the presence of truncation noise induced by a physical NLA. With realistic conditions of non-unit efficiency sources and detectors, a demonstrable improvement in channel transmission is achievable. Additionally, although parameters used in Fig.~\ref{fig:results3} were ambitious, it shows in principle the protocol can surpass the deterministic bound using a single quantum scissor. It is worth emphasizing that a significant strength of this protocol is that it works to correct the loss on the channel itself.  Therefore, it may be used on any optical field state regardless of the specific encoding of quantum information.

\section{Acknowledgements}
We thank Spyros Tserkis for useful discussions. This research was funded by the Australian Research Council Centre of Excellence for Quantum Computation and Communication Technology (Project No. CE110001027).

\section{Appendix}
Following the approach used in \cite{blandino2016channel,dias2017quantum}, we detail here the evolution of an input coherent state \(\ket{\alpha}_A\) through the error-correction protocol (Fig.~\ref{fig:error-correction1}). We include in this analysis losses in homodyne detection and single photon preparation and detection. 

Initially, the shared entanglement between Bob and Alice is of form:
\begin{equation}
\ket{\chi}_{RB} = \sqrt{1-\chi^2} \sum_{n=0}^\infty \chi^n\ket{n}_R \ket{n}_B .
\end{equation}

An arbitrary input state \(\ket{\psi}_A\) is mixed on a 50:50 beam splitter with mode \(R\) and dual homodyne detection is performed. Here \(\beta\) is detected, where 
\begin{equation}
\beta = X_-+iP_+
\end{equation}
with
\begin{align}
\hat{X}_- &= \hat{X}_A-\hat{X}_R 
\\
\hat{P}_+ &= \hat{P}_A+\hat{P}_R .
\end{align} 
 This measurement projects onto the eigenstate \cite{hofmann2000fidelity}
 \begin{equation}
 \ket{\beta}_{AR} = \frac{1}{\sqrt{\pi}}\sum_{n=0}^\infty \hat{D}_A(\beta)\ket{n}_A\ket{n}_R .
 \end{equation}
\begin{widetext}
With input state \(\ket{\psi}_A\), the output state conditioned on the measurement result \(\beta\) is therefore
\begin{align}
\ket{\psi(\beta)} &= _{AR}{\braket{\beta|\psi}}{_{A}}   \ket{\chi}_{RB}
\\
& = \frac{1}{\sqrt{\pi}} \sqrt{1-\chi^2} \sum_{n=0}^\infty \sum_{m=0}^\infty \bra{n}_A \bra{n}_R \chi^m \hat{D}_A \left(-\beta\right) \ket{\psi}_A \ket{m}_R \ket{m}_B .
\label{eq:output_cond}
\end{align}
The input state to the protocol is a coherent state \(\ket{\psi}_A=\ket{\alpha}_{A}\). Losses in the homodyne detection are modeled using a beam splitter of transmissivity \(\tau\) on modes  \(A\) and \(R\) before detection. This performs the following transformation on mode \(A\)
\begin{equation}
\ket{\psi}_A  \to \ket{\sqrt{\tau} \alpha}_A ,
\label{eq:A}
\end{equation}
and transforms mode \(R\) as
\begin{equation}
\ket{m}_R \to \hat{U}_{BS} \left[ \ket{m}_R \ket{0}_C \right]  = \sum_{k =0}^m \sqrt{\binom{m}{k}} \tau^{k/2} (1-\tau)^{(m-k)/2} \ket{k}_R \ket{m-k}_E .
\label{eq:R}
\end{equation}

Combining \eqref{eq:A}, \eqref{eq:R}, and \eqref{eq:output_cond}, the output state after detection of the measurement result \(\beta\) is:
\begin{equation}
\ket{\psi(\beta)} = \sqrt{\frac{1-\chi^2}{\pi}} e^{-|\sqrt{\tau}\alpha-\beta|^2/2} 
 \sum_{m=0}^\infty  \sum_{n=0}^m \chi^m  \sqrt{\binom{m}{n}} \tau^{n/2} (1-\tau)^{(m-n)/2}\frac{(\sqrt{\tau}\alpha-\beta)^n}{\sqrt{n!}}  \ket{m-n}_E  \ket{m}_B .
\end{equation}

Mode \(B\) then undergoes loss through channel attenuation modelled as a beam-splitter of transmission \(\eta\). We rescale this transmission as \(\eta=\frac{\nu}{\delta}\) to account for loss in the single photon detector \(D1\),
\begin{equation}
\ket{m}_B \to \hat{U}_{BS} \left[ \ket{m}_B \ket{0}_D \right]  = \sum_{j=0}^m \sqrt{\binom{m}{k}} \nu^{j/2} (1-\nu)^{(m-j)/2} \ket{j}_B \ket{m-j}_F .
\end{equation}

After channel attenuation, the state is
\begin{multline}
\ket{\psi(\beta)} = \sqrt{\frac{1-\chi^2}{\pi}} e^{-|\sqrt{\tau}\alpha-\beta|^2/2} 
\sum_{m=0}^\infty  \sum_{n=0}^m \chi^m  \sqrt{\binom{m}{n}} \tau^{n/2} (1-\tau)^{(m-n)/2}\frac{(\sqrt{\tau}\alpha-\beta)^n}{\sqrt{n!}}  \ket{m-n}_E  
\\
 \sum_{j=0}^m \sqrt{\binom{m}{j}} \nu^{j/2} (1-\nu)^{(m-j)/2} \ket{j}_B \ket{m-j}_F 
 \label{eq:epr_loss}
\end{multline}
 with modes \(E\) and \(F\) being loss modes, while mode \(B\) will be input into the NLA.

In the ideal case with no experimental inefficiencies, the NLA acts by combining mode \(B\) with a single photon in the form  \(\sqrt{\xi} \ket{1}_D \ket{0}_C + \sqrt{1-\xi}\ket{0}_D \ket{1}_C\) where the parameter \(\xi\) is related to the gain of the NLA by \(g=\sqrt{\frac{1-\xi}{\xi}}\). Modes \(B\) and \(D\) are then detected at \(D_1\) and \(D_2\) respectively. In the following, we model the realistic situation of imperfect single photon preparation and detection.

Preparation inefficiency of the single photon ancilla is modeled by a beam splitter with transmission \(\varepsilon\), mode \(G\) is a loss mode. 
\begin{equation}
\ket{\phi}_{NLA} \to \sqrt{\varepsilon}\ket{1}_D\ket{0}_G + \sqrt{1-\varepsilon}\ket{0}_D\ket{1}_G
\end{equation}   
The single photon ancilla \(\ket{\phi}_{NLA}\) then passes through the tunable beam splitter of transmission \(\xi\), mode \(C\) is the output mode. 
\begin{equation}
\ket{\phi}_{NLA} \to \sqrt{\varepsilon}\sqrt{\xi}\ket{1}_D \ket{0}_C \ket{0}_G + \sqrt{\varepsilon} \sqrt{1-\xi}\ket{0}_D \ket{1}_C \ket{0}_G + \sqrt{1-\varepsilon}\ket{0}_D\ket{0}_C\ket{1}_G
\end{equation}
The single photon ancilla \(\ket{\phi}_{NLA}\)  then undergoes loss \(\delta\) on mode \(D\) to model the loss in the single photon detector \(D2\), mode \(H\) is a loss mode. 
\begin{multline}
\ket{\phi}_{NLA} \to \sqrt{\varepsilon}\sqrt{\xi}  \sqrt{\delta} \ket{1}_D \ket{0}_C \ket{0}_H   \ket{0}_G    + \sqrt{\varepsilon} \sqrt{1-\xi}\ket{0}_D \ket{1}_C \ket{0}_H \ket{0}_G 
\\
+ \sqrt{\varepsilon}\sqrt{\xi}\sqrt{1-\delta} \ket{1}_H  \ket{0}_G \ket{0}_D \ket{0}_C + \sqrt{1-\varepsilon} \ket{0}_H \ket{1}_G\ket{0}_D \ket{0}_C  
\label{eq:phi}
\end{multline}
The single photon ancilla \eqref{eq:phi} is then combined with mode \(B\) \eqref{eq:epr_loss} on a 50:50 beam splitter and modes \(B\) and \(D\) are detected. A successful event is heralded when a single photon is detected at \(D1\) and none at \(D2\) or vice versa (\(\ket{0}_B \ket{1}_D\) or \(\ket{1}_D \ket{0}_D\)). 

The last step in the protocol is a displacement of the output mode by the measurement result \(\beta\) scaled by a classical gain \(\lambda\), given by \(\hat{D}_C\left( \lambda \beta \right) \).
 The entire, un-normalized output state of the protocol is:

\begin{multline}
\hat{\rho}_{out} 
= {\frac{1-\chi^2}{\pi}} e^{-|\sqrt{\tau}\alpha-\beta|^2}   \sum_{s=0}^\infty \sum_{r=0}^s  \left(1-\nu\right)^s  (1-\tau)^{r}   \chi^{2s} \hat{D}_C\left(\lambda \beta \right)  
\Bigg( \bigg\{   {\varepsilon}{\xi}  {\delta}  
  {\binom{s}{s-r}} \tau^{s-r}    \frac{\left(\left|\sqrt{\tau}\alpha-\beta\right|^2\right)^{s-r}}{{\left(s-r\right)!}}    + 
\\
+   
 \left[ {\nu} {\varepsilon}{\xi} \left(1-\delta \right)  +\left(1-\varepsilon \right) {\nu} \right]   \left(s+1\right) \chi^{2}  {\binom{s+1}{s+1-r}}   \tau^{\left(s+1-r\right)} 
 \frac{\left(\left|\sqrt{\tau}\alpha-\beta\right|^2\right)^{s+1-r}}{{\left(s+1-r\right)!}}     
 \bigg\} \ket{0}_C \bra{0}_C
\\
+{\varepsilon} \left( 1-\xi\right) {\nu}   
  \left(s+1\right) \chi^{2}  {\binom{s+1}{s+1-r}} \tau^{s+1-r}  \frac{\left(\left|\sqrt{\tau}\alpha-\beta\right|^2\right)^{s+1-r}}{{\left(s+1-r\right)!}}    \ket{1}_C \bra{1}_C
\\
+ {\varepsilon}\sqrt{\xi \delta \nu \tau \left( 1-\xi\right)}   (\sqrt{\tau}\alpha^*-\beta^*)
   \chi  {\binom{s}{s-r}} {\frac{s+1}{s+1-r}} \tau^{s-r} \frac{\left(\left|\sqrt{\tau}\alpha-\beta\right|^2\right)^{s-r}}{{\left(s-r\right)!}}        \ket{0}_C \bra{1}_C
\\
+    
{\varepsilon}\sqrt{\xi \delta \nu \tau \left( 1-\xi\right)}  (\sqrt{\tau}\alpha-\beta)
      \chi  {\binom{s}{s-r}} {\frac{s+1}{s+1-r}}   \tau^{s-r} \frac{\left(\left|\sqrt{\tau}\alpha-\beta\right|^2\right)^{s-r}}{{\left(s+1-r\right)!}}      \ket{1}_C \bra{0}_C
 \Bigg) \hat{D}_C^\dagger\left( \lambda \beta \right) .
\end{multline}

\end{widetext}
The variance of the output state was then calculated and averaged over the measurement outcome \(\beta\):
\begin{equation}
V = \int \braket{\hat{X}_C^2} \mathrm{d}^2\beta -\left(\int \braket{\hat{X}_C}\mathrm{d}^2 \beta\right)^2 .
\end{equation}

\end{document}